\documentclass{PoS}

\usepackage{amsmath,amssymb,amscd,amsfonts,graphicx,bbm}

\rightline{\sffamily RBRC-697, RIKEN-TH-119}\vspace*{-10mm}

\title{Phase structure of twisted Eguchi-Kawai model}

\ShortTitle{Phase structure of TEK model}

\author{\speaker{Tomomi Ishikawa}\\

        RIKEN BNL Research Center, Brookhaven National Laboratory,
	Upton, New York 11973, USA\\
        E-mail: \email{tomomi@quark.phy.bnl.gov}\vspace*{1mm}}

\author{Tatsuo Azeyanagi\\
        Department of Physics, Kyoto University,
	Kyoto 606-8502, Japan\vspace*{0.7mm}}

\author{Masanori Hanada\\
        Theoretical Physics Laboratory, RIKEN Nishina Center,
	Wako, Saitama 351-0198, Japan\vspace*{0.7mm}}

\author{Tomoyoshi Hirata\\
        Department of Physics, Kyoto University,
	Kyoto 606-8502, Japan\vspace*{10mm}}

\abstract{
We study the phase structure of the four-dimensional twisted Eguchi-Kawai
model using numerical simulations.
This model is an effective tool for studying $SU(N)$ gauge theory
in the large-$N$ limit and provides a nonperturbative formulation of 
the gauge theory on noncommutative spaces. 
Recently it was found that its $\mathbb{Z}_N^4$ symmetry,
which is crucial for the validity of this model,
can break spontaneously in the intermediate coupling region.
We investigate in detail the symmetry breaking point
from the weak coupling side.
Our simulation results show that the continuum limit of this model
cannot be taken.
}

\FullConference{The XXV International Symposium on Lattice Field Theory\\
                July 30 - August 4 2007\\
                Regensburg, Germany}

\begin{document}

\section{Introduction}

In 1982, Eguchi and Kawai introduced an important and interesting idea,
which is now called {\it Eguchi-Kawai equivalence} \cite{Eguchi:1982nm}.
Consider the $SU(N)$ gauge theory (YM) on a periodic
$D$-dimensional lattice with the Wilson plaquette action.
In the large-$N$ limit the space-time degrees of freedom can be neglected,
and the theory is then equivalent to a model defined on a single
hyper-cube, called the Eguchi-Kawai model (EK model).
This correspondence was shown by observing that the Schwinger-Dyson equations
for Wilson loops (loop equations) in both theories are the same.
Na\"ively, in the EK model the loop equations can have open Wilson
lines, which do not exist in the original gauge theory
due to gauge invariance.
Therefore we need to assume that the global $\mathbb{Z}_N^D$ symmetry
$U_\mu\to e^{i\theta_{\mu}}U_{\mu}$, which prohibits non-zero expectation
values of the open Wilson lines, is not broken spontaneously. 
However, soon after the discovery of the equivalence,
it was found that the symmetry is actually broken for $D>2$
in the weak coupling region \cite{Bhanot:1982sh}.
Although the naive EK equivalence does not hold,
several modifications have been proposed for this issue;
they are the ``quenched'' Eguchi-Kawai model (QEK model)
\cite{Bhanot:1982sh, Parisi:1982gp, Gross:1982at} and the
``twisted'' Eguchi-Kawai model (TEK model) \cite{GonzalezArroyo:1982ub}.
Historically, more work has been performed on the TEK model
because it is theoretically interesting and numerically more practical.
In addition, this model also describes gauge theories
on noncommutative spaces (NCYM) \cite{Aoki:1999vr, Ambjorn:1999ts}.

The TEK model is a matrix model defined by the partition function 
\begin{eqnarray}
Z_{TEK}=\int\prod_{\mu=1}^D dU_{\mu}\exp(-S_{TEK})
\end{eqnarray}
with the action
\begin{eqnarray}
S_{TEK}=-\beta N\sum_{\mu\neq\nu}Z_{\mu\nu}\mathrm{Tr}\
U_\mu U_\nu U_\mu^\dagger U_\nu^\dagger,\qquad \beta=1/g^2,
\label{action:TEK}
\end{eqnarray}
where $dU_{\mu}$ and $U_\mu\ (\mu=1,..,D)$ are Haar measure and link variables.
The phase factors $Z_{\mu\nu}$ are
\begin{equation}
Z_{\mu\nu}=\exp\left(2\pi i n_{\mu\nu}/N\right),\qquad
n_{\mu\nu}=-n_{\nu\mu}\in \mathbb{Z}_N.
\end{equation}
The classical solution $U^{(0)}_\mu=\Gamma_{\mu}$ satisfies the
't Hooft algebra
\begin{eqnarray}
\Gamma_{\mu}\Gamma_{\nu}=Z_{\nu\mu}\Gamma_{\nu}\Gamma_{\mu},
\label{'t Hooft algebra}
\end{eqnarray}
and is called ``twist-eater''.
The twist-eater guarantees existence of the $\mathbb{Z}_N^D$ symmetry 
in the weak coupling limit.
It is unclear whether or not the symmetry is unbroken in the intermediate
coupling region, as there is no guarantee the symmetry is preserved.
Numerical simulations performed during the early days of this model, however,
showed that the symmetry is unbroken throughout the whole coupling region.
This has encouraged the belief that the TEK model describes
the large-$N$ limit of $SU(N)$ gauge theory.

Surprisingly, some indications of $\mathbb{Z}_N^D$ symmetry breaking
were recently reported in several contexts concerning the TEK model
\cite{Ishikawa:2003, Bietenholz:2006cz, Teper:2006sp}.
In \cite{Teper:2006sp}, the $D=4$ TEK model with standard twist was
investigated up to $N=144$ and $\mathbb{Z}_N^4$ symmetry breaking phenomena
in the intermediate coupling region was observed by Monte-Carlo simulations.
The work presented in this article continues this investigation.
We concentrate on investigating the locations of the symmetry
breaking from the weak coupling side in the $(\beta, N)$ plane to determine
if the continuum limits of the TEK models can be taken as the YM and
the NCYM.

\section {Twist prescriptions}

In this study we treat the $D=4$ case.
Among the various types of twist possible, we apply:
\begin{eqnarray}
n_{\mu\nu}&=&\;\;\;L\ \epsilon_{\mu\nu}^{\rm sym},\qquad\;\, N=\;\;\; L^2
\qquad\mbox{(minimal symmetric twist, standard twist)},
\label{EQ:minimal symmetric twist}\\
n_{\mu\nu}&=&\;\;\; L\ \epsilon_{\mu\nu}^{\rm skew},\qquad N=\;\;\; L^2
\qquad\mbox{(minimal skew-diagonal twist)},
\label{EQ:minimal skew-diagonal twist}\\
n_{\mu\nu}&=&mL\ \epsilon_{\mu\nu}^{\rm skew},\qquad N=mL^2
\qquad\mbox{(generic skew-diagonal twist)},
\label{EQ:generic skew-diagonal twist}
\end{eqnarray}
where we define anti-symmetrization matrices as
\begin{equation}
\epsilon_{\mu\nu}^{\rm sym}=\left(\begin{array}{cccc}
 0 &  1 &  1 & 1\\
-1 &  0 &  1 & 1\\
-1 & -1 &  0 & 1\\
-1 & -1 & -1 & 0
\end{array}\right),\qquad
\epsilon_{\mu\nu}^{\rm skew}=\left(\begin{array}{cc|cc}
 0 &  1 &  0 & 0\\
-1 &  0 &  0 & 0\\
\hline
 0 &  0 &  0 & 1\\
 0 &  0 & -1 & 0
\end{array}\right).
\end{equation}
These twists represent $L^4$ lattices.
The symmetric and the skew-diagonal form can be transformed
into one another by an $SL(4, \mathbb{Z})$ transformation for the coordinates on
$\mathbb{T}^4$ \cite{van Baal:1985na}.
While these forms differ only by a coordinate transformation,
they can give different results except the weak coupling limit.
We note that the generic twist (\ref{EQ:generic skew-diagonal twist})
can be regarded as the gauge theory on $m$-coincident fuzzy $\mathbb{T}^4$.
(The minimal twists (\ref{EQ:minimal symmetric twist}) and
(\ref{EQ:minimal skew-diagonal twist})
are particular cases ($m=1$) of the generic twist.)

\section{Theoretical discussion for the $\mathbb{Z}_N^4$ symmetry breaking
         in the TEK model}
\label{SEC:theoretical}

As we mentioned in the introduction, the $\mathbb{Z}_N^4$ symmetry can be
broken in the intermediate coupling region.
In this section we give a theoretical discussion about this phenomena.

Here, we consider the first breaking point from the weak coupling limit
$\beta_c^L$.
We assume that $\mathbb{Z}_N^4$ symmetry breaking at this point
is a transition from the twist-eater phase $U_{\mu}=\Gamma_{\mu}$ to
the identity configuration phase $U_{\mu}=\mathbbm{1}_N$.\footnote{
Of course, the twist-eater only has $\mathbb{Z}_L^4$ symmetry for the
twists we apply. But we write it as $\mathbb{Z}_N^4$ in this article.}
For simplicity we consider a
$\mathbb{Z}_N^4\xrightarrow{\beta_c^L}\mathbb{Z}_N^0$ type transition here.
Of course we can treat
$\mathbb{Z}_N^4\xrightarrow{\beta_c^L}\mathbb{Z}_N^3\xrightarrow{\beta_c^L}
\mathbb{Z}_N^2\xrightarrow{\beta_c^L}\mathbb{Z}_N^1\xrightarrow{\beta_c^L}
\mathbb{Z}_N^0$ (cascade) type breaking,
but the obtained behavior is not different from the former type.
First, we focus on the classical energy difference between these
configurations.
The energy difference can be easily calculated from the action
(\ref{action:TEK}) as
\begin{eqnarray}
\Delta S
&=&S_{TEK}[U_{\mu}=\mathbbm{1}_N]-S_{TEK}[U_{\mu}=\Gamma_{\mu}]\nonumber\\
&=&\beta N^2\sum_{\mu\neq\nu}
   \left\{ 1-\cos\left(\frac{2\pi n_{\mu\nu}}{N}\right)\right\}
   \simeq 2\pi^2\beta\sum_{\mu\neq\nu}n_{\mu\nu}^2. 
\label{potential difference}
\end{eqnarray}
For the generic twist we have
\begin{equation}
\Delta S=
\begin{cases}
24\pi^2\beta m^2L^2\qquad & \mbox{(symmetric form)},\\
8\pi^2\beta m^2L^2\qquad  & \mbox{(skew-diagonal form)}.
\label{EQ:energy difference generic}
\end{cases}
\end{equation}
Note that the symmetric form is roughly three times more stable than
the skew-diagonal form if both twists have similar quantum fluctuations.
Thus the $\mathbb{Z}_N^4$ symmetry breaking for the skew-diagonal form
can occur at smaller $N$ than that for the symmetric form.

Going away from the weak coupling limit, the system experiences greater
quantum fluctuations.
Here, we naively expect that the $\mathbb{Z}_N^4$ symmetry is broken
if the fluctuation around the twist-eater configuration exceeds
the energy difference $\Delta S$.
Because the system describes $O(N^2)$ interacting gluons,
it is natural to assume that their quantum fluctuations provide
an $O(N^2)$ contribution to the effective action.
For the generic twist, the fluctuation is $O(m^2L^4)$.
Combining this with eq.~(\ref{EQ:energy difference generic}),
we can estimate the critical point $\beta_c^L$ as
\begin{equation}
\beta_c^L\sim L^2.
\label{EQ:beta_c estimation}
\end{equation}

Although the above discussion is quite crude,
the symmetry breaking behavior described by (\ref{EQ:beta_c estimation}) is
consistent with the numerical results discussed in the next section.

\section{Numerical simulations}

In this section we show the results of our numerical simulations for
the $\mathbb{Z}_N^4$ symmetry breaking phenomena.
In order to discuss the continuum and large-$N$ limits for this model,
we concentrate on the first breaking point from the weak coupling side.

\subsection{Simulation method}

In our simulation we use the pseudo-heatbath algorithm.
The algorithm is based on \cite{Fabricius:1984wp}, and in each sweep
over-relaxation is performed five times after multiplying
$SU(2)$ subgroup matrices.
The number of sweeps is $O(1000)$ for each $\beta$.
We scanned for the symmetry breaking point with a resolution of
$\Delta\beta=0.005$,
and thus quote $\pm 0.0025$ as the error due to the finite resolution. 
Note that the breaking points are ambiguous because the breakdown of
the $\mathbb{Z}_N^4$ symmetry is a first-order transition.
As an order parameter for detecting the symmetry breakdown,
we measure the expectation value of Polyakov lines
\begin{eqnarray}
P_\mu\equiv\left|\left\langle\frac{1}{N}\mathrm{Tr}\ U_\mu\right\rangle\right|.
\end{eqnarray}

\subsection{Simulation results}

\begin{figure}[tbp]
\begin{center}
\begin{tabular}{lr}
\begin{minipage}{70mm}
\hspace*{-4mm}
\includegraphics[scale=0.42, viewport = 0 0 510 470, clip]
                {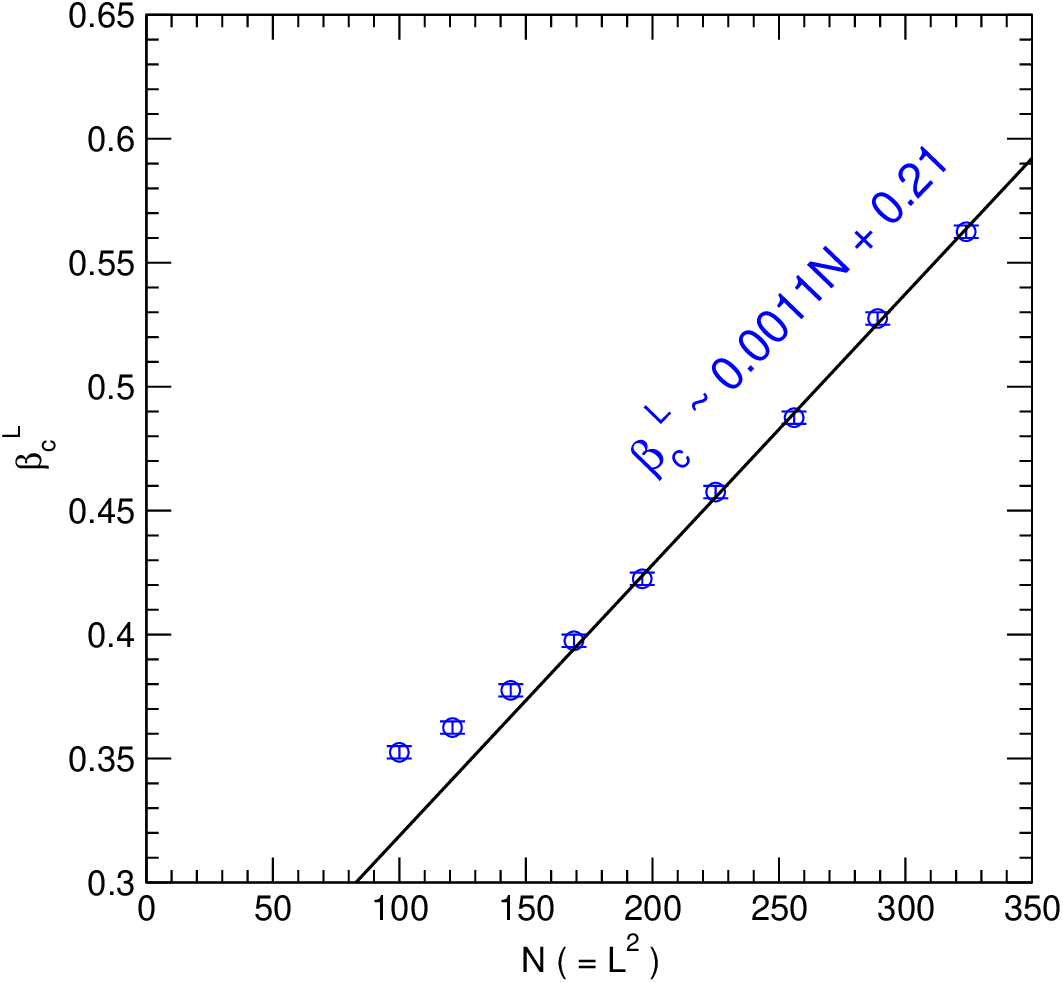}
\caption{$\beta_c^L$ versus $N$ (minimal symmetric twist).
         Fit line is eq.~(\protect\ref{EQ:bc_L minimal symmetric}),
         which is obtained using $N\geq169$ data.
         \vspace*{+5mm}}
\label{FIG:minimal symmetric twist}
\end{minipage}
&
\begin{minipage}{70mm}
\includegraphics[scale=0.42, viewport = 0 0 510 470, clip]
                {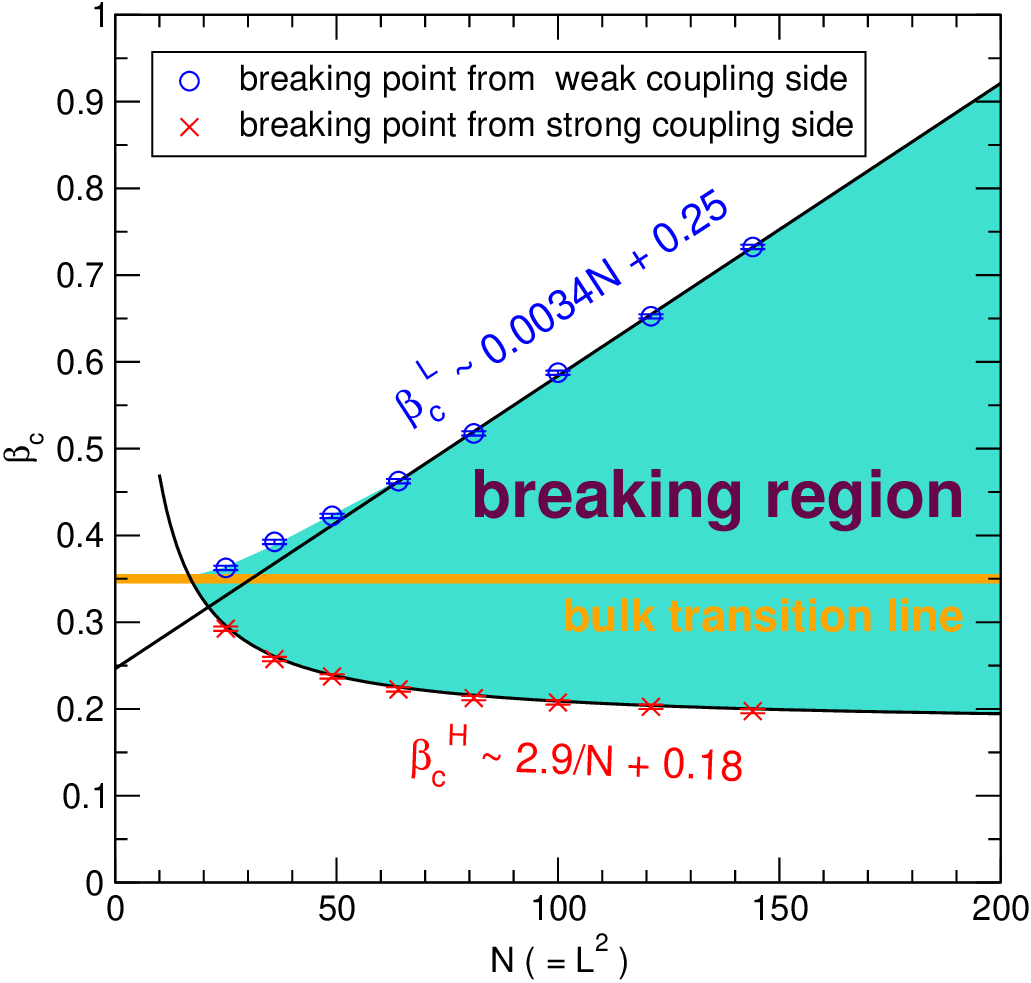}
\caption{$\beta_c^L$ and $\beta_c^H$ versus $N$ (minimal skew-diagonal twist).
         Fit lines are eqs.~(\protect\ref{EQ:bc_L minimal skew-diagonal})
         and (\protect\ref{EQ:bc_H minimal skew-diagonal}), which are
         obtained using $N\geq64$ and $N\geq25$ data, respectively.
         The $\mathbb{Z}_N^4$ symmetry is
         broken within the light blue shaded area.
         \vspace*{-5.1mm}}
\label{FIG:minimal skew-diagonal twist}
\end{minipage}
\end{tabular}
\end{center}
\end{figure}

First we consider the minimal twists.
Figures \ref{FIG:minimal symmetric twist} and
\ref{FIG:minimal skew-diagonal twist} show the critical lattice coupling
from the weak coupling side $\beta_c^L$ for the symmetric and skew-diagonal
twists, respectively.
For the minimal skew-diagonal twist we also observe the critical
lattice coupling from strong coupling side $\beta_c^H$.
We see that while the $\mathbb{Z}_N^4$ symmetry begins to break at $N=100$ for
the symmetric form, it is already violated at $N=25$ for the skew-diagonal
form, which is consistent with the theoretical considerations in section
\ref{SEC:theoretical}.
Additionally, we observe a clear linear dependence of $\beta_c^L$ on $N(=L^2)$:
\begin{eqnarray}
\beta_c^L&\sim&0.0011N+0.21 \qquad\mbox{(minimal symmetric twist)},
\label{EQ:bc_L minimal symmetric}\\
\beta_c^L&\sim&0.0034N+0.25 \qquad\mbox{(minimal skew-diagonal twist)},
\label{EQ:bc_L minimal skew-diagonal}
\end{eqnarray}
in the larger $N$ region.
This behavior was already obtained in the theoretical discussion.
Note that the coefficient of $N$ for the skew-diagonal twist is roughly
three times larger than that for symmetric twist, which is also
consistent with the theoretical analysis.
For $\beta_c^H$, we find the relation
\begin{equation}
\beta_c^H \sim 2.9/N+0.18 \qquad\mbox{(minimal skew-diagonal twist)}.
\label{EQ:bc_H minimal skew-diagonal}
\end{equation}
As $N$ is increased, $\beta_c^H$ approaches a point where the phase transition
$\mathbb{Z}_N^4\xrightarrow{\beta_c^H}\mathbb{Z}_N^3$ takes place
in the original EK model.

\begin{figure}[tbp]
\begin{center}
\begin{tabular}{lr}
\hspace*{-3mm}
\begin{minipage}{72mm}
\includegraphics[scale=0.42, viewport = 0 0 510 470, clip]
                {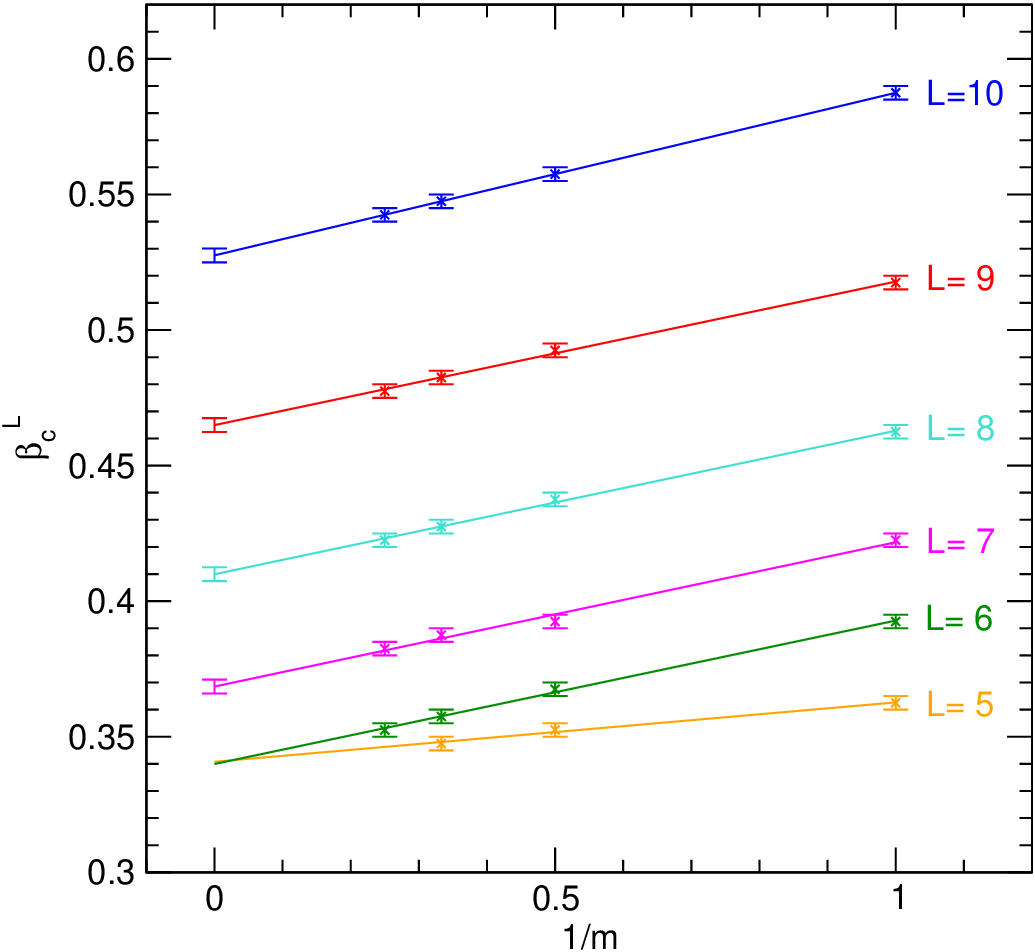}
\caption{$\beta_c^L$ versus $1/m$ for each $L$
         (generic skew-diagonal twist).
         $\beta_c^L$ for $m=\infty$ is obtained by linear extrapolation.
         \vspace*{-4.0mm}}
\label{FIG:bc_L_mall_skew1}
\end{minipage}
&
\begin{minipage}{72mm}
\includegraphics[scale=0.42, viewport = 0 0 510 470, clip]
                {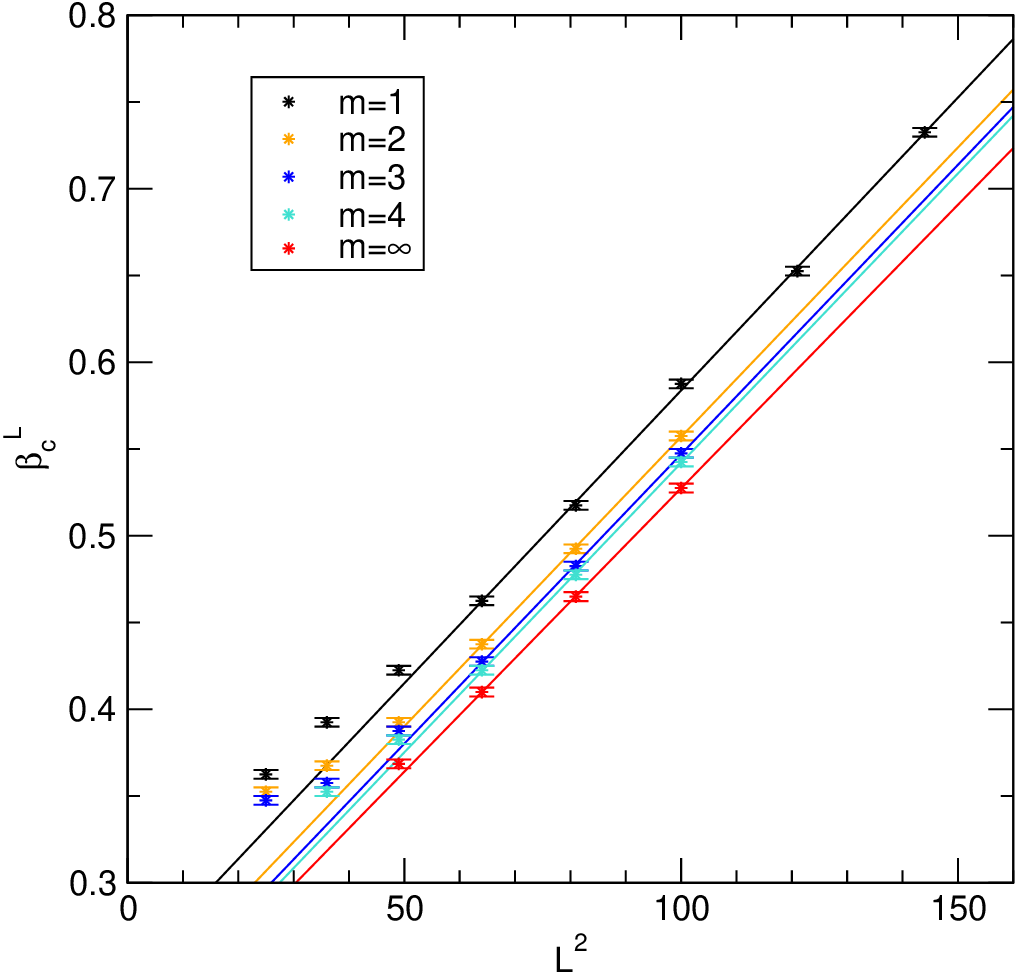}
\caption{$\beta_c^L$ versus $L^2$ for each $m$
         (generic skew-diagonal twist).
         We also include $m=\infty$ data, which is obtained
         by the extrapolation shown in figure \protect\ref{FIG:bc_L_mall_skew1}
         \vspace*{-5.7mm}}
\label{FIG:bc_L_mall_skew2}
\end{minipage}
\end{tabular}
\end{center}
\end{figure}

For the generic twist we use the skew-diagonal form because
$\mathbb{Z}_N^4$ symmetry breaking occurs at smaller $N$ than for
the symmetric form, which makes our investigation much easier.
We measure $\beta_c^L$ for this twist up to $m=4$.
The simulation results are plotted in figure \ref{FIG:bc_L_mall_skew1}.
From this figure we find that for each $L$, the $\beta_c^L$  show weak
$1/m$ behavior.
The points at $1/m=0$ are linearly extrapolated values.
($m=\infty$ means that an infinite number of fuzzy tori are superimposed.)
The behavior for $L=5$ is particularly interesting.
While $\mathbb{Z}_N^4$ symmetry breaking is observed for $m=1, 2,$ and $3$,
it is not seen for $m=4$ because $\beta_c^L$ reaches
a bulk transition point as $m$ is increased.
Figure \ref{FIG:bc_L_mall_skew2} represents the same data as
figure \ref{FIG:bc_L_mall_skew1}, but with $L^2$ as the horizontal axis.
As we have seen in the minimal case, the data for $L\geq8$ can be fitted
by a linear function in $L^2$ for each $m$.
From these figures, we discover that the data for $L\geq8$ are well
fitted globally by:
\begin{equation}
\beta_c^L\sim 0.0034L^2+\frac{0.060}{m}+0.19
 \qquad\mbox{(generic skew-diagonal twist)}.
\label{EQ:bc_L_generic_skew}
\end{equation}
The discussion in section \ref{SEC:theoretical}
did not predict the observed  dependence of $\beta_c^L$ on $1/m$.
While we do not have a clear reason for this phenomenon at present,
we suspect that it is related to collective modes of
the eigenvalues of the link variables.

\section{Continuum limit}

Although our simulation is restricted to the small $N$ region,
we may use our theoretical considerations to make statements about
the large-$N$ limit.
Thus the EK equivalence is valid only in the region
$\beta >\beta_c^L\sim N$, not only in the smaller $N$ region,
but also in the large-$N$ limit.

As is well known, the  one-loop perturbative calculation of the YM lattice
theory shows that its beta function  behaves as $\beta\sim\log a^{-1}$ near
the weak coupling limit, where $a$ is lattice spacing.
If we wish to make the TEK model correspond to the YM theory,
the scaling of the TEK model should obey that of the YM.
In the TEK model, the lattice size $L$ is related to $N$.
(For the generic twist, the relation is $N=mL^2$.)
Then the continuum limit of the YM system with fixed physical size
$l=aL$ can be obtained using the scaling
\begin{equation}
\beta\sim\log a^{-1}\sim\log N.
\label{scaling of the coupling in continuum limit}
\end{equation}
In order to obtain the large-$N$ limit with infinite volume,
we should keep $\beta$ lower than
eq.~(\ref{scaling of the coupling in continuum limit}).
Otherwise, the system would shrink to a point.
However, the simulation results obtained in this study show that
$\beta_c^L$ grows faster than the logarithm.
Therefore we conclude that EK equivalence breaks down
and the TEK model does not have YM as its continuum limit.

In the case of the NCYM, the beta function is essentially the same as that of
the YM theory at the one-loop level \cite{Minwalla:1999px},
and thus the scaling near the weak coupling limit
is $\beta\sim\log a^{-1}$.
But if we wish to make the TEK model correspond to the NCYM,
there is the constraint $a^2L={\rm fixed}$, which means that
we take a scheme in which the noncommutative parameter $\theta$ is fixed.
And then both the continuum limit and the infinite volume limit are
simultaneously taken (double scaling limit).
Regardless of the constraint,
by the nature of the logarithm scaling, the scaling for the NCYM
is the same as that of the ordinary YM
(eq.~\ref{scaling of the coupling in continuum limit}).
Therefore we conclude also that the TEK model does not have NCYM as
its continuum limit.

\section{Conclusions}

We carefully investigated the $\mathbb{Z}_N^4$ symmetry breaking
phenomena in the TEK model using Monte-Carlo simulation.
We found a clear linear dependence on $L^2$ for the symmetry breaking
point from the weak coupling side.
Regrettably, this means that the TEK models which use simple twists cannot
be made to correspond to either ordinary YM or NCYM in the continuum limit.

Finally, we mention the partial reduction \cite{Narayanan:2003fc},
which has been actively used in recent years.
\cite{Narayanan:2003fc} showed that the large-$N$ reduction is
valid above some critical physical size $l_c$.
This means that for a lattice size $L$ the reduction holds only below
some lattice coupling $\beta(L)$.
In order to take continuum limit we should avoid the bulk transition
at $\beta_c^B$, and thus there is a lower limit to the lattice size
$L_c(\beta)$ used for the reduction.
It is clever that the twist prescription is applied to this reduction 
\cite{GonzalezArroyo:2005dz}, and we believe it would be very efficient.

\acknowledgments

The numerical computations for this work were carried out in part at
the Yukawa Institute Computer Facility.
M.~H. is supported by Special Postdoctoral Researchers Program at RIKEN.
T.~H. would like to thank the Japan Society for the Promotion of Science
for financial support.

\end{document}